\documentclass[12pt,preprint]{aastex}
\usepackage{emulateapj5}

\newcommand{\pfrac}[2]{\left(\frac{#1}{#2}\right)}
\def\Mesz{M\'esz\'aros~}
\def\etal{{et al.~}}
\def\eps{\epsilon}

\slugcomment{Submitted to ApJL}

\shorttitle{Radio re-brightening of GRB afterglows}
\shortauthors{Li \& Song}

\begin{document}

\title{Late-Time Radio Re-Brightening of Gamma-Ray Burst Afterglows:\\ Evidence for Double-Sided Jets}

\author{Zhuo Li\altaffilmark{1} and L. M. Song\altaffilmark{1}}

\altaffiltext{1}{Particle Astrophysics Center, Institute of High
Energy Physics, CAS, Beijing 100039, China}

\begin{abstract}
The central engine of gamma-ray bursts (GRBs) is believed to eject
double-sided ultra-relativistic jets. For an observed GRB, one of
the twin jets should point toward us, and is responsible for the
prompt gamma-ray and subsequent afterglow emission. We consider in
this Letter the other receding jet, which will give rise to
late-time radio re-brightening (RRB) when it becomes
non-relativistic (NR) and radiative isotropic. The RRB peaks at a
time $5t_{NR}=2(E_{j,51}/n)^{1/3}$~yr after the GRB, where
$t_{NR}$ is the observed NR timescale for the preceding jet, $E_j$
is the jet energy and $n$ is the ambient medium density. The peak
flux is comparable to the preceding-jet emission at $t_{NR}$. We
expect the RRB of GRB 030329  1.7 yr after the burst with a flux
$\sim0.6$~mJy at 15~GHz. The cases of GRBs 970508 and 980703 have
also been discussed. The detection of RRB, which needs dense
monitoring campaign even a few years after a GRB, will be the
direct evidence for the existence of double-sided jets in GRBs,
and prove the black hole-disk system formation in the cores of
progenitors.
\end{abstract}


\section{Introduction}
A standard shock model of gamma-ray burst (GRB) afterglows has now
been well established, in which the GRB outflow (jet) drives a
relativistic expanding blast-wave that sweeps up and heats the GRB
ambient medium to produce the long-term X-ray, optical and radio
afterglow by synchrotron/inverse-Compton (IC) emission (see Zhang
\& \Mesz 2004 or Piran 2004 for recent reviews).

The remarkable supernova signature in GRB 030329 afterglow (Hjorth
\etal 2003) has confirmed that GRBs, at least the long-duration
ones, are originated from explosions of massive stars. Their
progenitors are believed to form black hole-disk systems in their
cores, and produce double-sided jets, which are responsible for
the GRB emission, along the spinning axis of the black hole (\Mesz
2002). For the observed hard GRB photons to escape freely,
avoiding electron/positron pair formation, the outflows are
reguired to be ultra-relativistic with Lorentz factor
$\Gamma\ga100$. Because of the relativistic beaming effect, the
GRB emission is confined in a narrow cone, and can only be
observed when the light of sight is within the cone.
Interestingly, it was this beaming effect that made many authors
expect the present of ``orphan afterglows'' (Rhoads 1997; Dalal,
Griest, \& Pruet 2002; Granot \etal 2002; Granot \& Loeb 2003)
which correspond to off-axis GRBs without gamma-rays associated,
but with afterglows detected when the shocks decelerate.

The previous works, except for Granot \& Loeb (2003), only
discussed the emission from the preceding jet (PJ), which is
pointing toward us, and neglected that from the other receding jet
(RJ), since it always points away from us. However, in this Letter
we discuss specially the emission from the RJ, which will emerge
when the RJ become non-relativistic (NR) and the radiation becomes
isotropic. Because of the light travel time delay, the RJ
contribution will overlay above the late-time, already decayed PJ
emission. In the NR phase the afterglow peak emission has moved to
radio frequencies, so the re-brightening from the RJ is only
expected in late-time radio afterglows. We predict the late-time
radio re-brightening (RRB) of GRB 030329 and discuss the long-term
radio observations for two GRBs 970508 and 980703. The observation
of RRBs will be particularly important due to its straightforward
implication of the existence of double-sided jets in GRBs.

\section{Emission from the receding jet}
Consider two collimated jets, that are ejected in opposite
directions from the central engine of a GRB. For an observed GRB,
the observer should be almost on the jet axis. We assume the two
jets have the same characteristics, such as the half opening angle
$\theta_j$, the initial Lorentz factor $\Gamma_0$, and the kinetic
energy $E_j/2$ (or with equivalent isotropic energy
$E_{iso}=2E_j/\theta_j^2$). Since most afterglow fits seem to
favor a uniform medium (Panaitescu \& Kumar 2002), we assume a
constant particle density $n$ for both jets.

In the above assumptions we have take the standard view of a
homogeneous jet with sharp edge (Rhoads 1999). Recently, a
structured jet model (\Mesz \etal 1998; Dai \& Gou 2001; Rossi
\etal 2002; Zhang \& \Mesz 2002) was proposed, but detailed fits
to afterglow data are still needed to determine whether it is
consistent with current observations (Granot \& Kumar 2003). Since
the standard jet model have been well successful in the data fits
(e.g., Panaitescu \& Kumar 2002), we only discuss homogeneous jets
here.

\subsection{Jet dynamics}
We first discuss the PJ. The PJ evolution can be divided into four
phases. (1) Initially, $\Gamma\gg1/\theta_j$, the transverse size
of the jet is larger than that of the causally connected region,
therefore the jet evolves as if it was a conical section of a
spherical relativistic blast wave. The PJ undergoes first a
coasting phase with $\Gamma=\Gamma_0$.  (2) After the jet-induced
shock sweeps up enough medium, the jet kinetic energy is mostly
transferred into the shocked medium, and the jet begins to
decelerate significantly. This occurs at the deceleration radius
$r_d=3E_{iso}/(4\pi n\Gamma_0^2m_pc^2)$ at an observer time
$t_d=r_d/(2\Gamma_0^2c)$, where a relation $dr\approx2\Gamma^2cdt$
has been taken for surperluminal motion. At the deceleration
phase, the dynamics could be well described by the Blandford \&
McKee (1976) self-similar solution, where $\Gamma\propto
r^{-3/2}\propto t^{-3/8}$. (3) As the jet  continues to
decelerate, the Lorentz factor drops to $1/\theta_j$ at a time
$t_j=t_d(\Gamma_0\theta_j)^{8/3}$, corresponding to a radius
$r_j=r_d(\Gamma_0\theta_j)^{2/3}$. From the time on the transverse
size of causally connected regions exceeds that of the jet,
therefore the sphere approximation breaks down, and the jet starts
expanding sideways. The dynamical evolution in this stage depends
on the degree of sideways expansion. If the lateral velocity in
the comoving frame equals to the local sound speed, the jet will
spread quickly with the opening angle increasing and $\Gamma$
dropping exponentially (Rhoads 1999). At this stage, the radius
hardly increases and may be regarded as a constant $r\approx r_j$,
and $\Gamma\propto t^{-1/2}$. (4) Finally the jet goes into NR
phase with $\Gamma\sim1$ at $t_{NR}=t_j/\theta_j^2$, where the
evolution can be well described by Sedov-Taylor solution
$\beta\propto r^{-3/2}\propto t^{-3/5}$, with
$\beta=(1-1/\Gamma^2)^{1/2}$. In summary, the evolution is:
\begin{equation}
\begin{array}{lll}
\mbox{coasting:} &  t<t_d & \Gamma=\Gamma_0,~   r\propto t \\
\mbox{sphere:} &   t_d<t<t_j & \Gamma\propto t^{-3/8},~  r\propto t^{1/4}  \\
\mbox{spreading:}  &  t_j<t<t_{NR} & \Gamma\propto t^{-1/2},~  r\approx r_j  \\
\mbox{NR:} & t>t_{NR} & \beta\propto t^{-3/5}~(\Gamma\sim1),~
r\propto t^{2/5}
\end{array}
\end{equation}
Using the above simplified dynamical relation, the NR time and
radius, defined as where $\Gamma=1$, are able to be calculated as
\begin{equation}\label{eq:nr}
t_{NR}=\frac1{2c}\pfrac{3E_j}{2\pi m_pc^2n}^{1/3},~ r_{NR}\approx
r_j=\pfrac{3E_j}{2\pi m_pc^2n}^{1/3}.
\end{equation}
They are independent of $\Gamma_0$ and $\theta_j$. With typical
values $E_j=10^{51}E_{j,51}$~ergs and $n=1$~cm$^{-3}$ (e.g., Frail
\etal 2001; Panaitescu \& Kumar 2001), we have
\begin{equation}\label{eq:tnr}
t_{NR}=130\pfrac{E_{j,51}}{n}^{1/3}{\rm d}.
\end{equation}
Note that this $t_{NR}$ refers to the PJ.

As for the RJ, it also transits into NR phase at a radius $r_{NR}$.
However, due to the light travel time the observer time epoch is delayed
by a time $2r_{NR}/c$, so the observed NR time for the RJ is
\begin{equation}\label{eq:rjtnr}
t_{NR}^{RJ}=t_{NR}+\frac {2r_{NR}} c=5t_{NR},
\end{equation}
5 times that of the PJ\footnote{It can be shown that the factor 5
is also valid for the wind case where the external density
decreases as the square of the distance from the source (Dai \& Lu
1998; Chevalier \& Li 2000).}.

There are uncertainties about the lateral spreading of jets, and
it might be that little spreading occurs, as shown in some
numerical simulations (e.g., Cannizzo \etal 2004). If we assume
the extreme case that the lateral velocity is zero, then the jet
remains conical geometry and continues to evolve as if it was a
conical section of a spherical relativistic blast wave after
$\Gamma<1/\theta_j$. In this case, the NR time for the PJ is
$t_{NR}=(1/2)(3E_j/2\pi
m_pc^5n\theta_j^2)^{1/3}=610(E_{j,51}/n\theta_{j,-1}^2)^{1/3}$~d,
with $\theta_{j,-1}=\theta_j/10^{-1}$, and the corresponding
radius is $r_{NR}=2ct_{NR}$, therefore for the RJ, the NR time is
$t_{NR}^{RJ}=t_{NR}+2r_{NR}/c=5t_{NR}=8.4(E_{j,51}/n\theta_{j,-1}^2)^{1/3}$~yr.
These may be regarded as  upper limits, while the values in eqs.
(\ref{eq:nr}-\ref{eq:rjtnr}) for the fastest spreading cases as
lower limits.

\subsection{Non-relativistic phase emission}
For the detailed discussion of afterglow emission in NR phase, the readers
might  refer to Frail, Waxman \& Kulkarni (2000), Livio \& Waxman
(2000) and Waxman (2004), we here only give brief introduction.

Consider also the PJ emission first. After $t_{NR}$, the evolution
is described by the spherical NR self-similar evolution, with the
PJ radius given by $r=r_{NR}(t/t_{NR})^{2/5}$, and the swept-up
particle number is $N_e\simeq(4/3)\pi nr^3$. Simply assume that
the shocked material forms a thin and uniform shell, with the
width $\Delta=r/\eta\ll r$ and the post-shock thermal energy
density $U=E_j/(4\pi r^2\Delta)$. The electron and magnetic-field
energy densities are assumed to be constant fraction $\eps_e$ and
$\eps_B$, respectively, of the post-shock thermal energy density
$U$. The electrons are believed to be accelerated to form a power
law distribution, $dn_e/d\gamma_e\propto\gamma_e^{-p}$ for
$\gamma_e\ge\gamma_m$. Based on the above assumptions, it can be
found that the magnetic field strength and the minimum electron
Lorentz factor evolve as $B\propto t^{-3/5}$ and $\gamma_m\propto
t^{-6/5}$, respectively.

The accelerated electrons will give rise to synchrotron
emission. The characteristic synchrotron frequency of electrons
with $\gamma_m$ is $\nu_m=eB\gamma_e^2/(2\pi m_ec)$, i.e.,
\begin{equation}
\nu_m=0.3\eps_{e,-1}^2\eps_{B,-1}^{1/2}n^{1/2} \pfrac
t{t_{NR}}^{-3}\mbox{GHz},
\end{equation}
where $\eps_{e,-1}=\eps_e/10^{-1}$, and
$\eps_{B,-1}=\eps_B/10^{-1}$. The peak specific luminosity is
$L_{\nu_m}=N_eP/\nu_m$ where
$P=(4/3)\sigma_Tc\gamma_m^2(B^2/8\pi)$ is the synchrotron power of
an electron with $\gamma_m$. The observed radio frequency is
usually beyond $\nu_m$ at NR stage, the specific luminosity is
therefore $L_\nu=L_{\nu_m}(\nu/\nu_m)^{(1-p)/2}\propto
t^{21/10-3p/2}$, and given by
\begin{eqnarray}
L_\nu=4\times10^{30}\eps_{e,-1}\eps_{B,-1}^{3/4}n^{3/4}E_{j,51}\nonumber\\
\pfrac\nu{10~ {\rm GHz}}^{-1/2}\pfrac t{t_{NR}}^{-9/10}\mbox{ergs
s$^{-1}$Hz$^{-1}$},
\end{eqnarray}
where $p=2$ has been taken. The NR light curve flattens, compared
with the decay, $\propto t^{-p}$, in the spreading phase (Rhoads
1999; Sari, Piran \& Halpern 1999).

As for the RJ, when it is still relativistic the emission is
always beamed away from us and hence invisible. At the time
$t_{NR}^{RJ}$ after the GRB, the RJ becomes non-relativistic and
the emission becomes isotropic. The observed RJ luminosity at this
point should be comparable to that of the PJ at $t_{NR}$, because
of the same characteristics of the PJ and RJ. If neglect the NR
jet velocity with respect to the light speed, the time delay
between the emission from the PJ and RJ could be roughly fixed as
$2r/c\approx2r_{NR}/c=4t_{NR}$ (for $r$ is not $\gg r_{NR}$). The
RJ specific luminosity, then, could be written as, for $t\geq
t_{NR}^{RJ}$,
\begin{equation}\label{eq:LnuRJ}
L_\nu^{RJ}(t)\simeq L_\nu(t-4t_{NR})\propto \pfrac
{t-4t_{NR}}{t_{NR}}^{-9/10}.
\end{equation}
A schematic plot is given in figure 1 for both PJ and RJ radio
light curves. After rapid increase in flux, the RJ emission
reaches the peak at around $t_{NR}^{RJ}$, and then decreases as  a
delayed light curve of the NR-phase PJ, but showing a steeper
decline in the logarithmic plot. Note that the transition to NR
phase might be gradual, and the jet edge might be not sharp,
therefore both these can smooth the RRB feature on the plot.

\section{Observation}

\subsection{GRB 970508 and GRB 980703}
As discussed above, if a GRB can be monitored for a long time in radio bands,
e.g., a few years, the RRB may be able to be
detected. So far, there are two GRBs that have been reported
in the literatures with radio monitoring longer than one year:
GRB 970508 and GRB 980703 (Frail, Waxman \& Kulkarni 2000;
Berger, Kulkarni \& Frail 2001; Frail \etal 2003).

Frail, Waxman \& Kulkarni (2000) report  an extensive monitoring
of the radio afterglow of GRB 970508, lasting 450 days after the
burst. In the data analysis they found that the spectral and
temporal radio behavior indicate a transition to NR expansion at
$t_{NR}\approx100$~d, therefore, we should expect the RRB at
around day 500. However, this is unfortunately behind the last
observation of this burst, so we might have missed that apparent
phenomena.

GRB 980703 was monitored even longer, up to $\sim10^3$~d, as shown
in figure 1 of Frail \etal (2003). Two obvious flattening appear
in the radio light curves: while the late-time transition to a
constant flux is thought to be the host galaxy contribution, the
earlier flattening at $\approx40$~d is attributed to the
transition into NR expansion (Berger, Kulkarni \& Frail 2001).
Modelling the $\Gamma$ evolution has also inferred a similar value
of $t_{NR}=30-50$~d (Frail \etal 2003). Therefore, the expected
RRB should arise at $\sim200$~d, and then after another time
interval of 200~d, around 400~d from the burst, the radio flux
should decline to a value similar to just before the RRB. At first
glance of figure 1 of Frail \etal (2003), the  radio light curves,
somewhat more extensive at 8.5 and 4.8~GHz, seem to show only
decays without any re-brightening. However, we notice that there
is no observation data collected between 210~d and $\sim400$~d,
therefore the RRB may be missed once again due to the lack of
observation.

\subsection{Prediction for GRB 030329}
GRB 030329 is the nearest cosmological burst ($z=0.1685$), and has
a bright afterglow at all wavelengths. Berger \etal (2003) have
reported its bright radio afterglow up to $\sim70$~d after the
burst. Since no radio flattening (due to transition to NR
evolution) appears yet\footnote{Although the X-ray afterglow shows
a flattening around day 37 (Tiengo \etal 2004), we believe this is
not relevant to the relativistic to NR transition of the radio
jet, because of no simultaneous radio flattening (Berger \etal
2003), and also conflicted with the still superluminal expansion
in the radio angular size measurement (Taylor \etal 2004). The
X-ray flattening may need other explanations, e.g. an IC
component; NR transition for the narrow jet in two-component model
(Berger \etal 2003).}, we should calculate $t_{NR}$ using $E_j/n$
derived from available data.

The early breaks ($\sim0.5$~d) in the R band (Price \etal 2003)
and X-ray (Tiengo \etal 2004) light curves infer a small jet
opening angle, leading to a small jet-corrected energy release in
GRB 030329, which is more than one order of magnitude below the
average value around which most GRBs are narrowly clustered (Frail
\etal 2001; Panaitescu \& Kumar 2001). Two models have been
proposed to solve this problem: a refreshed-shock model with 10
times more energy injected later (Granot, Nakar \& Piran 2003);
and a two-component model with most energy in a wider jet
component (Berger \etal 2003). Both models have the same result:
the kinetic energy corresponding to the late-time radio afterglow
is typical for all GRBs. Based mainly on the early multi-frequency
data, a typical value for medium density, $n=1$~cm$^{-3}$, was
also derived by Willingale \etal (2004). Therefore, the estimated
value of $t_{NR}$ should be typical as in eq. (\ref{eq:tnr}),
hence $t_{NR}\sim130$~d. In fact, the radio angular size
measurement of this burst shows relativistic expanding 83~d after
the burst (Taylor \etal 2004), consistent with the estimation
here.

In addition, Berger \etal (2003) have inferred
$E_{j,51}/n=0.14\nu_{c,13}^{-1/2}$ from a snapshot spectrum at
$t_j\approx10$~d, where $\nu_c=\nu_c/10^{13}$~Hz is the cooling
frequency. This parameter value leads to a too early NR phase,
$t_{NR}\approx68$~d (from eq. \ref{eq:tnr}), conflicted with the
observed superluminal expansion (Taylor \etal 2004). However, in
the parameter derivation, the assumption of jet break time at
$t_j\approx10$~d might be incorrect, because the millimeter
observation shows that the light curves have already steepened at
$\approx5$~d at both 100 and 250~GHz (Sheth \etal 2003), implying
that $t_j\la5$~d. Moreover, the angular size evolution can also
give constraints to the model parameters. For GRB 030329, Granot,
Ramirez-Ruiz \& Loeb (2004) derive $E_{j,51}/n\sim0.8$ (see also
Oren, Nakar \& Piran 2004) from the observation by Taylar \etal
(2004). This yields a later transition at $t_{NR}\sim120$~d.

We will take the more reasonable value $t_{NR}\approx120$~d (It
should be noticed that this might still be the lower limit since
the most rapid sideways spreading has been assumed), and hence the
expected RRB should arise at $\sim1.7$~yr after GRB 030329, with
the fluxes comparable to that at the time $t_{NR}$. Assuming the
same temporal slope for late time, an extrapolation of the
available radio light curves (Berger \etal 2003) to $\sim120$~d
gives a flux $\sim0.6$~mJy at 15~GHz, or $\sim0.3$~mJy at 44~GHz.
Dense monitoring campaign around $\sim1.7$~yr after GRB 030329 is
required to obtain a well-observed RRB profile.

\section{Summary and discussion}
In this work we suggest that RRBs are common for GRBs with
double-sided jets, which come from the RJ transition to NR
evolution. If we assume the same properties for both PJ and RJ and
also for the ambient medium on both sides, the RRB would arise at
a time $t_{NR}^{RJ}=5t_{NR}=1.8(E_{j,51}/n)^{1/3}$~yr after the
burst, with a flux comparable to that at the time $t_{NR}$. After
$t_{NR}^{RJ}$, the afterglow behaves as a delayed emission of that
behind $t_{NR}$, with the time lag of $4t_{NR}$ (eq.
\ref{eq:LnuRJ}). Unfortunately, no radio data could be collected
when RRBs occur for GRB 970508 and GRB 980703, two longest
observed GRBs so far. We suppose the RRB of GRB 030329 around
1.7~yr after the burst with $\sim0.6$~(0.3)~mJy at 15~(44)~GHz,
and urge dense monitoring campaign during that time. It should be
noticed that for weak jet-spreading, the RRB might be more delayed
since $t_{NR}^{RJ}$ is much larger (see in the end of \S 2.1).

There has been growing evidence for collimated jets in GRBs over
the past several years, which is  coming mainly from observations
of achromatic breaks in the afterglow light curves (e.g., Kulkarni
et al. 1999; Stanek et al. 1999). However, there are still other
explanations for the light curve breaks, for example, the
transition from the relativistic  to  NR phase of the blast wave
at a few days due to highly dense medium (Dai \& Lu 1999; Wang,
Dai \& Lu 2000); the effects of IC scattering flattening or
steepening the light curves (Wei \& Lu 2000); a sudden drop in the
external density (Kumar \& Panaitescu 2000); a break in the energy
spectrum of radiating electrons (Li \& Chevalier 2001). On the
other hand, though black hole-disk system with twin jets is
generally assumed in GRBs, magnetized, rapidly rotating neutron
stars remain contenders, and an off-center dipole could lead to a
one-sided jet. Since RRBs are only associated with collimated,
double-sided, relativistic outflows, the detection of RRBs would
provide straightforward evidence of double-sided jets in GRBs, and
prove the black hole-disk system formation in the cores of
progenitors. This make RRB observations significantly important.

\acknowledgments

Z. Li thanks R. F. Shen for discussions, and L. J. Gou for
comments. This work was supported by the National 973 Project and
the Special Funds for Major State Basic Research Projects.

\begin{figure}[ht]
\epsscale{.7} \plotone{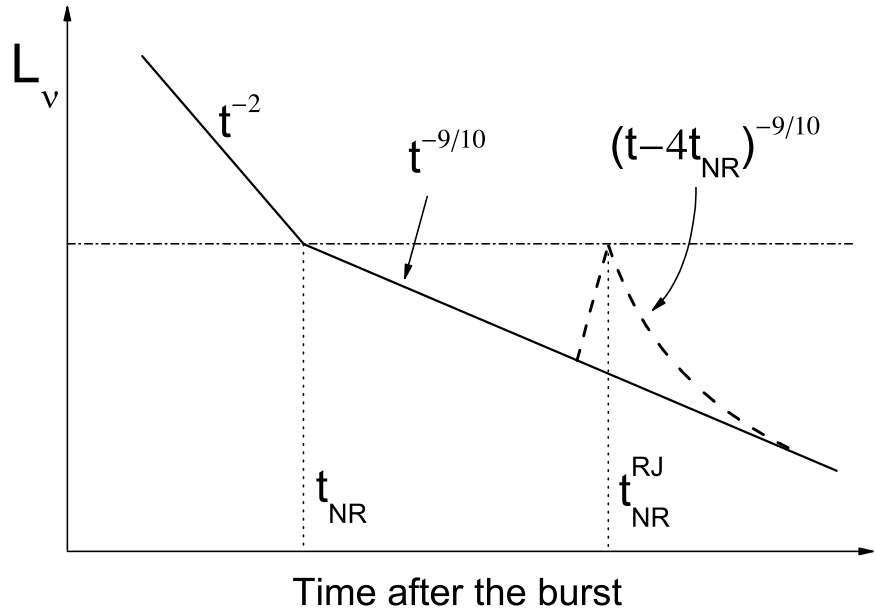} \caption{The schematic log-log plot
for the radio light curve at $\nu>\nu_m$. The solid and dashed
lines correspond to the contribution from the PJ and RJ,
respectively. The time scalings are marked, assuming $p=2$. After
a steep decline in the spreading phase, the PJ light curve flatten
at $t_{NR}$. The RJ emission rapidly increases to a peak around
$t_{NR}^{RJ}=5t_{NR}$, and then declines following the way of the
PJ unless a time delay of $4t_{NR}$, but exhibiting steep light
curve in the log-log plot. The dashed-dot line indicates
$L_{\nu}(t_{NR})=L_\nu^{RJ}(t_{NR}^{RJ})$.}
\end{figure}


\begin{thebibliography}{}

\bibitem[Berger et al.(2003)]{2003Natur.426..154B} Berger, E., et al.\
2003, \nat, 426, 154

\bibitem[Berger, Kulkarni, \& Frail(2001)]{2001ApJ...560..652B} Berger, E.,
Kulkarni, S.~R., \& Frail, D.~A.\ 2001, \apj, 560, 652

\bibitem[Cannizzo, Gehrels, \& Vishniac(2004)]{2004ApJ...601..380C}
Cannizzo, J.~K., Gehrels, N., \& Vishniac, E.~T.\ 2004, \apj, 601, 380

\bibitem[]{422} Chevalier, R. A., \& Li, Z.-Y.\ 2000, ApJ, 536, 195

\bibitem[Dai \& Gou(2001)]{2001ApJ...552...72D} Dai, Z.~G.~\& Gou, L.~J.\
2001, \apj, 552, 72

\bibitem[]{427} Dai, Z.~G.~\& Lu, T.\ 1998, \mnras, 298, 87

\bibitem[Dai \& Lu(1999)]{1999ApJ...519L.155D} Dai, Z.~G.~\& Lu, T.\ 1999,
\apjl, 519, L155

\bibitem[Dalal, Griest, \& Pruet(2002)]{2002ApJ...564..209D} Dalal, N.,
Griest, K., \& Pruet, J.\ 2002, \apj, 564, 209

\bibitem[Frail et al.(2001)]{2001ApJ...562L..55F} Frail, D.~A., et al.\
2001, \apjl, 562, L55

\bibitem[Frail et al.(2003)]{2003ApJ...590..992F} Frail, D.~A., et al.\
2003, \apj, 590, 992

\bibitem[Frail, Waxman, \& Kulkarni(2000)]{2000ApJ...537..191F} Frail,
D.~A., Waxman, E., \& Kulkarni, S.~R.\ 2000, \apj, 537, 191

\bibitem[]{444} Granot, J.~\& Kumar, P.\ 2003, \apj, 591, 1086

\bibitem[Granot \& Loeb(2003)]{2003ApJ...593L..81G} Granot, J.~\& Loeb, A.\
2003, \apjl, 593, L81

\bibitem[Granot, Nakar, \& Piran(2003)]{2003Natur.426..138G} Granot, J.,
Nakar, E., \& Piran, T.\ 2003, \nat, 426, 138

\bibitem[Granot, Panaitescu, Kumar, \& Woosley(2002)]{2002ApJ...570L..61G}
Granot, J., Panaitescu, A., Kumar, P., \& Woosley, S.~E.\ 2002,
\apjl, 570, L61

\bibitem[]{453} Granot, J., Ramirez-Ruiz, E., \& Loeb, A.\ 2004, (astro-ph/0407182)

\bibitem[]{456} Hjorth, J., \etal \ 2003, \nat, 423, 847

\bibitem[Kulkarni et al.(1999)]{1999Natur.398..389K} Kulkarni, S.~R., et
al.\ 1999, \nat, 398, 389

\bibitem[Kumar \& Panaitescu(2000)]{2000ApJ...541L..51K} Kumar, P.~\&
Panaitescu, A.\ 2000, \apjl, 541, L51

\bibitem[Li \& Chevalier(2001)]{2001ApJ...551..940L} Li, Z.-Y.~\& Chevalier,
R.~A.\ 2001, \apj, 551, 940

\bibitem[Livio \& Waxman(2000)]{2000ApJ...538..187L} Livio, M.~\& Waxman,
E.\ 2000, \apj, 538, 187

\bibitem[]{470} M{\' e}sz{\' a}ros, P. 2002, \araa,  40, 137

\bibitem[Meszaros, Rees, \& Wijers(1998)]{1998ApJ...499..301M} M{\' e}sz{\' a}ros,
P., Rees, M.~J., \& Wijers, R.~A.~M.~J.\ 1998, \apj, 499, 301

\bibitem[]{475} Oren, Y., Nakar, E., \& Piran, T.\ 2004, preprint (astro-ph/0406277)

\bibitem[Panaitescu \& Kumar(2001)]{2001ApJ...560L..49P} Panaitescu, A.~\&
Kumar, P.\ 2001, \apjl, 560, L49

\bibitem[Panaitescu \& Kumar(2002)]{2002ApJ...571..779P} Panaitescu, A.~\&
Kumar, P.\ 2002, \apj, 571, 779

\bibitem[]{483} Piran, T. 2004, Rev. Mod. Phys., accepted (astro-ph/0405503)

\bibitem[Price et al.(2003)]{2003Natur.423..844P} Price, P.~A., et al.\
2003, \nat, 423, 844

\bibitem[]{488}{{Rhoads}, J.~E.}\ 1997, \apj, 487, L1

\bibitem[Rhoads(1999)]{1999ApJ...525..737R} Rhoads, J.~E.\ 1999, \apj, 525,
737

\bibitem[Rossi, Lazzati, \& Rees(2002)]{2002MNRAS.332..945R} Rossi, E.,
Lazzati, D., \& Rees, M.~J.\ 2002, \mnras, 332, 945

\bibitem[Sari, Piran, \& Halpern(1999)]{1999ApJ...519L..17S} Sari, R.,
Piran, T., \& Halpern, J.~P.\ 1999, \apjl, 519, L17

\bibitem[Sheth et al.(2003)]{2003ApJ...595L..33S} Sheth, K., Frail, D.~A.,
White, S., Das, M., Bertoldi, F., Walter, F., Kulkarni, S.~R., \&
Berger, E.\ 2003, \apjl, 595, L33

\bibitem[Stanek et al.(1999)]{1999ApJ...522L..39S} Stanek, K.~Z.,
Garnavich, P.~M., Kaluzny, J., Pych, W., \& Thompson, I.\ 1999,
\apjl, 522, L39

\bibitem[Taylor, Frail, Berger, \& Kulkarni(2004)]{2004ApJ...609L...1T}
Taylor, G.~B., Frail, D.~A., Berger, E., \& Kulkarni, S.~R.\ 2004,
\apjl, 609, L1

\bibitem[Tiengo et al.(2004)]{2004A&A...423..861T} Tiengo, A., Mereghetti,
S., Ghisellini, G., Tavecchio, F., \& Ghirlanda, G.\ 2004, \aap,
423, 861

\bibitem[Wang, Dai, \& Lu(2000)]{2000MNRAS.317..170W} Wang, X.~Y., Dai,
Z.~G., \& Lu, T.\ 2000, \mnras, 317, 170

\bibitem[Waxman(2004)]{2004ApJ...602..886W} Waxman, E.\ 2004, \apj, 602,
886

\bibitem[Wei \& Lu(2000)]{2000A&A...360L..13W} Wei, D.~M.~\& Lu, T.\ 2000,
\aap, 360, L13

\bibitem[Willingale et al.(2004)]{2004MNRAS.349...31W} Willingale, R.,
Osborne, J.~P., O'Brien, P.~T., Ward, M.~J., Levan, A., \& Page,
K.~L.\ 2004, \mnras, 349, 31

\bibitem[Zhang \& M{\' e}sz{\' a}ros(2002)]{2002ApJ...571..876Z} Zhang,
B.~\& M{\' e}sz{\' a}ros, P.\ 2002, \apj, 571, 876

\bibitem[]{541} Zhang, B.~\& M{\' e}sz{\' a}ros, P.\ 2004, Int. J. Mod. Phys. A, 19, 2385

\end{thebibliography}
\end{document}